\documentclass[%
 reprint,
superscriptaddress,
 amsmath,amssymb,
 aps,
prb,
twocolumn
]{revtex4-2}

\usepackage{graphicx,bm,braket,color,hyperref,comment}
\usepackage{fancybox}
\usepackage{ulem}
\hypersetup{colorlinks=true,citecolor=blue,linkcolor=blue,urlcolor=blue}

\begin{document}


\title{Three-stage Phase Transitions and Field-induced Phases in CeCoSi: A Landau Theory
}

\author{Takayuki Ishitobi}
\affiliation{Advanced Science Research Center, Japan Atomic Energy Agency, Tokai, Ibaraki 319-1195, Japan}

\date{\today}

\begin{abstract}
We investigate both the nonmagnetic and magnetic ordered phases of CeCoSi using Landau theory. Our analysis predicts three successive phase transitions at zero magnetic field. A quadrupole order parameter that emerges below $T_0=12$ K acts as a weak symmetry-breaking field on the antiferromagnetic ordering below $T_{\rm N}=9.4$ K, leading to two-stage magnetic transitions. 
In the higher-temperature antiferromagnetic phase within the range $T_{\rm s2}=8~{\rm K}<T<T_{\rm N}$, an out-of-plane component of the antiferromagnetic moment may or may not be present. If present, magnetic fields applied along the [100] and [110] directions induce additional magnetic phases. 
\end{abstract}

\pacs{Valid PACS appear here}
\maketitle

Identifying order parameters is a fundamental issue in condensed matter physics, yet it remains challenging for certain compounds such as ${\rm URu_2Si_2}$ \cite{Mydosh2014-kf}. Such difficulty often arises in systems where no experimental probe directly couples to the order parameter. In such cases, it is essential to constrain the candidate order parameters within a broad theoretical framework independent of their microscopic details. Landau theory provides a robust foundation for analyzing second-order phase transitions \cite{LandauLifshitz_StatisticalPhysics}. By incorporating crystal symmetries and external fields, one can classify possible order parameters and construct phase diagrams without relying on specific microscopic mechanisms.

CeCoSi is a recent example of a system that exhibits unresolved ordering phenomena. It crystallizes in the CeFeSi-type structure (space group $P4/nmm$, No. 129) and exhibits nonmagnetic ordering at $T_0 = 12$ K and antiferromagnetic ordering at $T_{\rm N} = 9.4$ K \cite{Tanida2019-ar}. The nonmagnetic ordered phase is attributed to a $(Q_{zx}, Q_{yz})$-type ferroquadrupolar order, at least from a symmetry perspective \cite{Matsumura2022-ky, Manago2023-yk}, while the antiferromagnetic phase is believed to be characterized by an in-plane magnetic moment with wave vector ${\bm q}={\bm 0}$ \cite{Nikitin2020-zu}. 
Since an ordering of localized quadrupole is seemingly incompatible with the crystalline electric field scheme \cite{Nikitin2020-zu}, the quadrupolar order is considered unconventional. 
Despite extensive theoretical \cite{Yatsushiro2020-gt, Yatsushiro2020-iy, Yatsushiro2022-qf, Yamada2024-sm} and experimental 
\cite{Lengyel2013-tn, Tanida2019-ar, Nikitin2020-zu, Kawamura2020-in, Matsumura2022-ky, Manago2023-yk, Kimura2023-sg, Kanda2024-vh} studies, the nature of its order parameters and the properties of the field-induced phases remain elusive. 

\begin{figure*}[t!] 
\centering 
\includegraphics[width=0.85\textwidth]{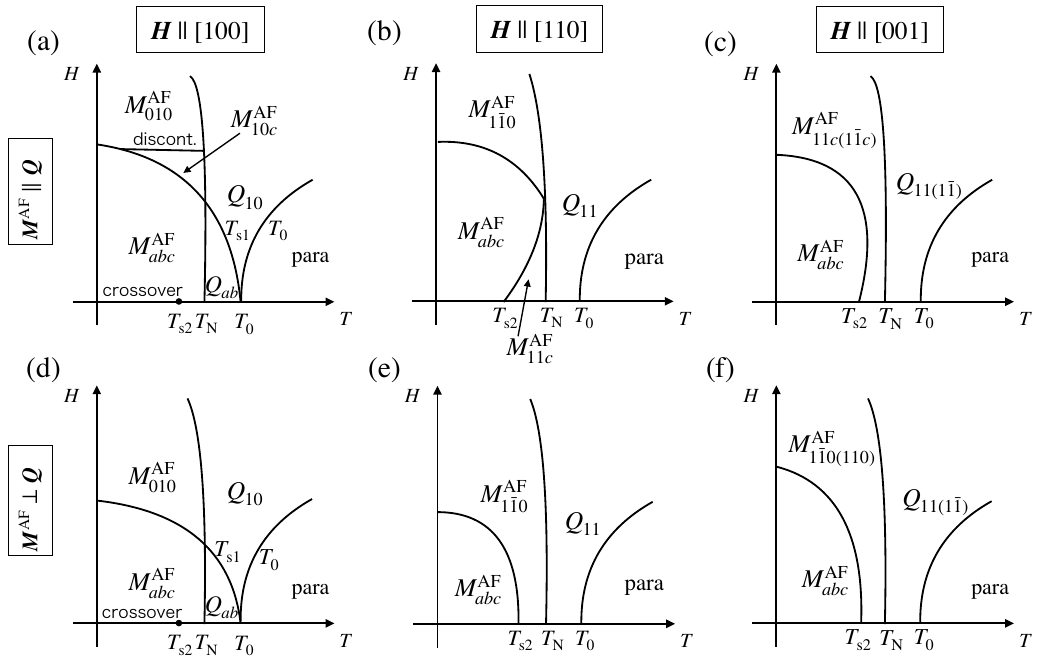} 
\caption{Schematic temperature--magnetic field phase diagrams for (a)--(c) ${\bm M}^{\rm AF} \parallel {\bm Q}$ and (d)--(f) ${\bm M}^{\rm AF} \perp {\bm Q}$. (a, d) ${\bm H} \parallel [100]$; (b, e) ${\bm H} \parallel [110]$; and (c, f) ${\bm H} \parallel [001]$. In the figures, the order parameter $Q_{ab}$ denotes $aQ_{zx}+bQ_{yz}$, and $M^{\rm AF}_{abc}$ represents $aM_x^{\rm AF}+bM_y^{\rm AF}+cM_z^{\rm AF}$. Phases characterized by $Q_{ab}$ and $M^{\rm AF}_{abc}$ are triclinic, while other ordered phases are monoclinic.} 
\label{fig:PhaseDiagrams}
\end{figure*}

Although the microscopic detail of the order parameter remains unclear, its symmetry has been constrained by experimental observations, including lattice distortions detected via X-ray diffraction (XRD) \cite{Matsumura2022-ky}, nuclear magnetic resonance (NMR) \cite{Manago2023-yk}, and the anisotropy of the temperature--magnetic field phase diagrams~\cite{Matsumura2022-ky, Hidaka2025-ky}. In particular, XRD measurements reveal Bragg peak splitting below $T_0$~\cite{Matsumura2022-ky}, and the magnetic phase diagram shows strong anisotropy, with a field-induced phase within the temperature range $T_{\rm s1}<T<T_0$ appearing only under magnetic fields applied along the [100] direction. These findings suggest a $(Q_{zx}, Q_{yz})$-type ferroquadrupolar order. However, the compatibility of this scenario with the magnetic phase diagrams and its implications for magnetic ordering at lower temperatures remain insufficiently understood.
In this letter, we perform a Landau theoretical analysis and construct temperature--magnetic field phase diagrams together with the corresponding order parameters, as summarized in Fig. \ref{fig:PhaseDiagrams}.

We first summarize the existing experimental data to derive reasonable assumptions in Landau theory. The present study is primarily based on the interpretation of XRD measurements~\cite{Matsumura2022-ky}, and thus, we begin with the implications of these observations. As mentioned earlier, the XRD experiments reveal the splitting of Bragg reflections below the transition temperature $T_0$. Specifically, the $(207)$ and $(118)$ reflections split into two and three peaks, respectively, below $T_0 = 12$~K, and further into four peaks below $T_{\rm s2} = 8$~K. While Ref.~\cite{Matsumura2022-ky} concluded that the crystal structure becomes triclinic below $T_0$, such an interpretation would imply the splitting into four peaks already below $T_0$, leaving no room for further crystalline symmetry lowering at $T_{\rm s2}$. A more consistent interpretation is that the crystal structure becomes monoclinic below $T_0$ and triclinic below $T_{\rm s2}$. 

The data under magnetic fields are also crucial to determine possible order parameters. Under a magnetic field along the $[010]$ direction, the $(207)$ reflection does not split in the temperature range $T_{\rm s1} < T < T_0$, leading to the conclusion that the structure in this phase is orthorhombic or tetragonal~\cite{Matsumura2022-ky}. However, this interpretation does not adequately consider the possibility that the orientation of the order parameter may differ between zero field and finite field, as we will demonstrate later. If the order parameter corresponds to a ferroquadrupole of the $\pm Q_{yz}$ type, which induces a monoclinic distortion, the tilt of the crystal axes lies within the $(207)$ reflection plane. In such a case, the absence of splitting in the $(207)$ reflection is naturally explained.

Furthermore, in the presence of a $[010]$ magnetic field, two of the four split reflections observed below $T_{\rm s2}$ are significantly weakened, and only two peaks remain visible for $H \geq 2$~T. This suggests that no structural transition occurs below $T_{\rm s1}$ under these conditions. For $[\bar{1}10]$ magnetic fields, the splitting of the $(118)$ reflection is suppressed, and a single peak is observed for $H \geq 4$~T. This observation is consistent with an order parameter of the form $\pm(Q_{zx} - Q_{yz})$, which induces a tilt of the crystal axes within the $(118)$ reflection plane, thereby leading to no splitting.
Based on these experimental data and their interpretation, we construct a Landau free energy and determine the signs of the coefficients of its various terms.

We now perform a Landau theoretical analysis, beginning with the couplings between order parameters. Based on experimental evidences, we consider the ferroquadrupolar order parameter ${\bm Q}=(Q_{zx}, Q_{yz})\equiv Q(\cos \phi_{Q}, \sin \phi_{Q})$ and the antiferromagnetic one ${\bm M}^{\rm AF} = (M^{\rm AF}_x, M^{\rm AF}_y)\equiv M^{\rm AF}(\cos \phi_{M}, \sin \phi_{M})$.  
For notational simplicity, we denote order parameters $aQ_{zx}+bQ_{yz}$ and $aM^{\rm AF}_x+bM^{\rm AF}_y+cM^{\rm AF}_z$, as $Q_{ab}$ and $M^{\rm AF}_{abc}$, respectively. Phases with nonzero $|a| \neq |b|$ have triclinic crystalline symmetry, while other ordered phases are monoclinic.

Reflecting the in-plane anisotropy ${\rm Re}[(x + iy)^4] = (x^2 - y^2)^2 - (2xy)^2$ of the $D_{4h}$ point group, the free energy includes the following anisotropic term for quadrupoles: 
\begin{align}
\delta F^Q &=
c^Q[(Q_{zx}^2 - Q_{yz}^2)^2 - (2Q_{zx}Q_{yz})^2] \nonumber \\
&= c^Q Q^4\cos(4\phi_{Q}).
\label{eq:Quad}
\end{align}
The sign of the coupling constant $c^Q$ determines whether the stable order is of $Q_{10}$-type ($\phi_{Q}=n\pi/2$) or $Q_{11}$-type ($\phi_{Q}=\pi/4+n\pi/2$) with $n$ being an integer. The $Q_{11}$-type with $c^Q>0$ is consistent with the XRD experiments \cite{Matsumura2022-ky}.  
Note that both signs of $c^Q$ results in monoclinic phase preserving (100)- or (110)-type mirror symmetry, as discussed in Ref. \cite{Hidaka2025-ky}.

Symmetry-allowed biquadratic couplings between ${\bm Q}$ and ${\bm M}^{\rm AF}$ take the form
\begin{align}
\delta F^{MQ} &= c^{MQ}_{x^2-y^2}(Q_{zx}^2-Q_{yz}^2)[(M^{\rm AF}_{x})^2-(M^{\rm AF}_{y})^2] \nonumber \\
&~~+4c^{MQ}_{xy}Q_{zx}Q_{yz}M^{\rm AF}_{x}M^{\rm AF}_{y}  \nonumber \\
& = \sum_{\pm} c^{MQ}_{\pm} Q^2(M^{\rm AF})^2 \cos(2\phi_{Q} \mp 2\phi_{M}).
\label{eq:mq_biq}
\end{align}
In the last line, the coefficient $c^{MQ}_\pm$ is defined by $c^{MQ}_\pm = (c^{MQ}_{x^2-y^2} \pm c^{MQ}_{xy})/2$, where the $c^{MQ}_{-}$ term represents tetragonal anisotropy, while the $c^{MQ}_+$ term is in-plane isotropic.  
This coupling is quadratic in ${\bm M}^{\rm AF}$, and thus dictates that the second-order phase transition from the quadrupolar to the antiferromagnetic phase must result in either ${\bm M}^{\rm AF} \parallel {\bm Q}$ ($c^{MQ}_{x^2-y^2}<0$ for $c^Q<0$ and $c^{MQ}_{xy}<0$ for $c^Q>0$) or ${\bm M}^{\rm AF} \perp {\bm Q}$ ($c^{MQ}_{x^2-y^2}>0$ for $c^Q<0$ and $c^{MQ}_{xy}>0$ for $c^Q>0$). Hereafter, we use ${\bm M}^{\rm AF} \parallel {\bm Q}$ to indicate that ${\bm M}^{\rm AF}$ and ${\bm Q}$ are parallel or antiparallel. In both cases, the resulting monoclinic phase preserves the crystal structure. Consequently, a third phase transition must occur at zero field to account for the structural phase transition to the triclinic phase observed at $T_{\rm s2}=8$ K by XRD measurement \cite{Matsumura2022-ky}. Such a phase transition arises when the sign of the tetragonal anisotropic coupling constant $c^M$ for the magnetic moment differs from that for the quadrupole moment. The relevant coupling term is
\begin{align}
\delta F^{M} &=
c^M \{ [(M^{\rm AF}_{x})^2 - (M^{\rm AF}_{y})^2]^2 - [2M^{\rm AF}_{x}M^{\rm AF}_{y}]^2 \} \nonumber \\
&=c^M (M^{\rm AF})^4\cos(4\phi_{M}),
\label{eq:AFM}
\end{align}
which is obtained by replacing ${\bm Q}$ with ${\bm M}^{\rm AF}$ in Eq. (\ref{eq:Quad}). When $c^Q>0$ and $c^M<0$, Eq.~(\ref{eq:AFM}) is minimized by a configuration with ${\bm M}^{\rm AF} \parallel {\bm x}$ or ${\bm y}$. 
In this case,  $\delta F^{MQ}$ in Eq.~(\ref{eq:mq_biq}) and $\delta F^M$ in Eq.~(\ref{eq:AFM}) compete. Note that $\delta F^{MQ}$ is second order in ${\bm M}^{\rm AF}$ and therefore dominates over the fourth order term $\delta F^{M}$ at the antiferromagnetic ordering temperature $T_{\rm N}$. At lower temperatures, $\delta F^{M}$ becomes dominant over $\delta F^{MQ}$, resulting in the third phase transition. These three-stage phase transitions can be understood as a consequence of the quadrupolar order parameter acting as a weak symmetry-breaking field for the antiferromagnetic transitions, analogous to the role of the antiferromagnetic order parameter in the superconducting transitions of ${\rm UPt_3}$~\cite{Sauls1994-dn}. The negative sign of $c^M$ is consistent with the crystalline electric field scheme suggested by high-field magnetization measurements \cite{Kanda2024-vh}. Since the signs of the coupling constants in Eq.~(\ref{eq:mq_biq}) cannot be determined from existing experiments, we consider both ${\bm M}^{\rm AF} \parallel {\bm Q}$ and ${\bm M}^{\rm AF} \perp {\bm Q}$ scenarios for the intermediate phase in $T_{\rm s2} < T < T_{\rm N}$. 

The quadrupolar order hybridizes the in-plane and out-of-plane components of ${\bm M}^{\rm AF}$ through a coupling
\begin{align}
\delta F^{QM_z} &=
c^{QM_z}M_z^{\rm AF}(M_x^{\rm AF}Q_{zx} + M_y^{\rm AF}Q_{yz}) \nonumber \\
&=c^{QM_z}M_z^{\rm AF}M^{\rm AF}Q \cos(\phi_{Q}-\phi_{M})
\label{eq:mmq}
\end{align}
with typically $c^{QM_z}<0$. Consequently, a ${\bm M}^{\rm AF} \parallel {\bm Q}$ component induces an out-of-plane moment $M_z^{\rm AF}$.  
This coupling and the resulting canting of magnetic moments have recently been discussed in the contexts of noncoplanar magnetic orders and the anomalous Hall effects in materials such as DyAuGe~\cite{Kurumaji2025-cc} and URhSn~\cite{Ishitobi2025-nn}. Although this coupling is isotropic within the $c$ plane and therefore has only a minor impact on the anisotropy of the phase diagrams, detecting the out-of-plane canting of the magnetic moment would provide a means to experimentally distinguish whether ${\bm M}^{\rm AF} \parallel {\bm Q}$ or ${\bm M}^{\rm AF} \perp {\bm Q}$ is realized.

Next, we examine the coupling to an external magnetic field ${\bm H} = (H_x, H_y, H_z)\equiv (H_\perp \cos \phi_H, H_\perp \sin \phi_H, H_z)$. Substituting ${\bm M}^{\rm AF} \rightarrow {\bm H}$ into Eq.~(\ref{eq:mq_biq}) yields the coupling between ${\bm Q}$ and ${\bm H}$:
\begin{align}
\delta F^{HQ} &= 
c^{HQ}_{x^2-y^2}(Q_{zx}^2-Q_{yz}^2)(H_x^2-H_y^2) \nonumber \\
&~~+ 4c^{HQ}_{xy}Q_{zx}Q_{yz}H_xH_y \nonumber \\
&=\sum_{\pm} c^{HQ}_{\pm} Q^2H_\perp^2 \cos(2\phi_{Q} \mp 2\phi_{H}),
\label{eq:hhqq}
\end{align}
where $c^{HQ}_{\pm} = (c^{HQ}_{x^2-y^2} \pm c^{HQ}_{xy})/2$, with $c^{HQ}_-$ and $c^{HQ}_+$ representing tetragonal anisotropy and in-plane isotropy, respectively. 
This coupling is quadratic in ${\bm Q}$ and constrains the orientation of ${\bm Q}$ below $T_0$ to be either ${\bm Q} \parallel {\bm H}$ or  ${\bm Q} \perp {\bm H}$, as discussed in Ref. \cite{Hidaka2025-ky}. As summarized earlier, experimental results of XRD measurements suggest that $Q_{01}$-type and $Q_{1\bar{1}}$-type orders are realized in a magnetic field along $[010]$ and $[\bar{1}10]$, respectively \cite{Matsumura2022-ky}. In both cases, ${\bm Q} \parallel {\bm H}$ is realized.  
Since the field-induced configuration $Q_{10}$ under a [100] field differs from zero-field configuration $Q_{11(1\bar{1})}$, it manifests as a field-induced phase. In contrast, a field along [110] preserves the zero-field configuration, and no field-induced phase emerges, in agreement with experiments \cite{Matsumura2022-ky, Hidaka2025-ky}. In terms of symmetry-breaking field, a magnetic field along the $[100]$ direction acts as a weak symmetry-breaking field for the $Q_{11(1\bar{1})}$-type order parameter, while a field along the $[110]$ direction does not.

The coupling between ${\bm M}^{\rm AF}$ and ${\bm H}$, analogous to Eq.~(\ref{eq:mq_biq}), is given by
\begin{align}
\delta F^{HM} &= c^{HM}_{x^2-y^2}[(M^{\rm AF}_x)^2 - (M^{\rm AF}_y)^2](H_x^2-H_y^2) \nonumber \\
&~~+ 4c^{HM}_{xy}M^{\rm AF}_xM^{\rm AF}_yH_xH_y \nonumber \\
&=\sum_{\pm} c^{HM}_{\pm} H^2(M^{\rm AF})^2 \cos(2\phi_{\rm H} \mp 2\phi_{M}),
\label{eq:mh_biq}
\end{align}
where analogous to Eqs.~(\ref{eq:mq_biq}) and (\ref{eq:hhqq}), $c^{HM}_\pm$ is defined by $c^{HM}_\pm = (c^{HM}_{x^2-y^2} \pm c^{HM}_{xy})/2$, where $c^{HM}_-$ and $c^{HM}_+$ terms represent tetragonal anisotropy and in-plane isotropy, respectively. 
The coupling constants $c^{HM}_{x^2-y^2}$ and $c^{HM}_{xy}$ are typically positive, stabilizing ${\bm M}^{\rm AF} \perp {\bm H}$. At low magnetic fields, the magnetic anisotropy is governed by Eqs.~(\ref{eq:mq_biq}) and (\ref{eq:AFM}), while Eq. (\ref{eq:mh_biq}) becomes dominant at high fields.

We now discuss the temperature--field phase diagrams and the corresponding order parameters. Let us first summarize the phase transitions at zero magnetic fields. 
At $T = T_0$, a $Q_{11(1\bar{1})}$-type ferroquadrupolar order occurs, where its orientation is determined by $\delta F^Q$ in Eq.~(\ref{eq:Quad}) with $c^Q > 0$. 
At $T = T_{\rm N}$, antiferromagnetic ordering emerges, with its orientation governed by $\delta F^{MQ}$ in Eq.~(\ref{eq:mq_biq}). Depending on the sign of the coupling coefficient $c^{MQ}_{xy}$ in Eq.~(\ref{eq:mq_biq}), either $M^{\rm AF}_{11c}$ or $M^{\rm AF}_{1\bar{1}0}$ is realized, corresponding to ${\bm M}^{\rm AF} \parallel {\bm Q}$ or ${\bm M}^{\rm AF} \perp {\bm Q}$, respectively. The sign of $c^{MQ}_{xy}$ cannot be determined from existing experiments. If ${\bm M}^{\rm AF} \parallel {\bm Q}$, the magnetic moment acquires a finite $c$-axis component. Upon further lowering the temperature, the fourth-order anisotropy term $\delta F^M$ in Eq.~(\ref{eq:AFM}) becomes dominant over the biquadratic coupling $\delta F^{MQ}$ in Eq.~(\ref{eq:mq_biq}), leading to an in-plane reorientation of the magnetic and quadrupole moments. This reorientation corresponds to the transition at $T = T_{\rm s2}$ into the $M^{\rm AF}_{abc}$ phase. 
In the following, we discuss the temperature--field phase diagrams for ${\bm H}\parallel$ [100], [110], and [001].

For ${\bm H} \parallel [100]$, $\delta F^{HQ}$ in Eq.~(\ref{eq:hhqq}) determines the orientation of the ferroquadrupole order parameter at $T_0$, instead of $\delta F^Q$ in Eq.~(\ref{eq:Quad}) at zero fields. As a result, a field-induced $Q_{10}$-type phase emerges at $T_0$, followed by a low-temperature phase below $T_{\rm s1}$ with an order parameter of the form $Q_{ab}$. As the latter order lowers the crystal symmetry to triclinic, the antiferromagnetic order parameter below $T_{\rm N}$ marks as $M^{\rm AF}_{abc}$. Thus, the zero-field transition into $M^{\rm AF}_{abc}$ at $T_{\rm s2}$ becomes a crossover in finite [100] fields. According to $\delta F^{HM}$ in Eq.~(\ref{eq:mh_biq}), the low-temperature high-field state stabilizes a configuration with $M_{010(c)}^{\rm AF} \perp {\bm H}$. This matches the zero-field order at $T_{\rm N}$ only if ${\bm M}^{\rm AF} \perp {\bm Q}$, while a mismatch occurs for ${\bm M}^{\rm AF} \parallel {\bm Q}$, indicating the emergence of a field-induced phase. In the latter case, $M^{\rm AF}_{10c}$ is favored at low fields, where the magnetic moment is canted toward the $c$-axis due to the coupling $\delta F^{QM_z}$ in Eq.~(\ref{eq:mmq}). At high fields, a phase $M_{010}$ with the antiferromagnetic moment perpendicular to the magnetic field is stable because of Eq.~(\ref{eq:mh_biq}), where the magnetic moment remains untilted because it is perpendicular to the quadrupole.  
The schematic phase diagrams are shown in Fig.~\ref{fig:PhaseDiagrams}(a, d), where the phase boundaries are drawn so that the symmetry of each phase lowers progressively with decreasing temperature. In both cases, the distinct combinations of presence or absence of $Q_{01}$ and ${\bm M}^{\rm AF}$ in the order parameters lead to four distinct phases that intersect at a tetracritical point. 

For ${\bm H} \parallel [110]$, both $\delta F^{HQ}$ and $\delta F^Q$ favor the $Q_{11}$-type order, and no field-induced quadrupolar ordered phase appears. At high fields and low temperatures, $M^{\rm AF}_{1\bar{1}0(c)} \perp {\bm H}$ is stabilized due to $\delta F^{HM}$ in Eq.~(\ref{eq:mh_biq}). This matches the zero-field order at $T_{\rm N}$ only for ${\bm M}^{\rm AF} \perp {\bm Q}$, yielding the simple phase diagram shown in Fig.~\ref{fig:PhaseDiagrams}(e). If instead ${\bm M}^{\rm AF} \parallel {\bm Q} \parallel {\bm H}$ is realized, field-induced successive phase transitions can occur to the $M^{\rm AF}_{1\bar{1}0}$-type phase and the intermediate $M^{\rm AF}_{abc}$-type phase, yielding a tetracritical point, as in Fig.~\ref{fig:PhaseDiagrams}(b). Alternatively, a direct first-order transition from the $M^{\rm AF}_{11c}$ to $M^{\rm AF}_{1\bar{1}0}$ may appear with two tricritical points connected by the first-order transition line.

A field along [001] does not affect in-plane anisotropy and thus has only a minor effect on the phase diagram. However, the coupling 
$\delta F^{QMH_z}=c^{QMH_z}H_zMQ \cos (\phi_Q-\phi_{FM})$, which is obtained by replacing $M_z^{\rm AF}$ with $H_z$ and ${\bm M}^{\rm AF}$ with magnetization ${\bm M}=M(\cos \phi_{FM}, \sin \phi_{FM})$ in Eq.~(\ref{eq:mmq}), induces an in-plane magnetization component ${\bm M} \parallel {\bm Q}$. The emergence of the magnetization suppresses the transition temperature of the parallel component ${\bm M}^{\rm AF} \parallel {\bm M}$ and relatively enhances that of ${\bm M}^{\rm AF} \perp {\bm M}$. The resulting phase diagrams are shown in Figs.~\ref{fig:PhaseDiagrams}(c, f). In Fig.~\ref{fig:PhaseDiagrams}(c), $T_{\rm s2}$ and $T_{\rm N}$ are assumed not to cross. If they do cross, the resulting topology of the phase diagram matches that of Fig.~\ref{fig:PhaseDiagrams}(b) under a [110] field.

We now discuss the consistency of our theoretical results with existing experiments and possible future experimental approaches to further validate our predictions. In this study, we showed that successive antiferromagnetic transitions are required to account for the structural phase transition observed at $T_{\rm s2} = 8$~K in zero magnetic fields. This scenario is consistent with the experimental observation that the structural transition temperature $T_{\rm s2}$ observed via XRD~\cite{Matsumura2022-ky} differs from the antiferromagnetic transition temperature $T_{\rm N} = 9.4$~K identified through specific heat and magnetization measurements~\cite{Tanida2019-ar, Hidaka2025-ky}. Moreover, under a magnetic field along the $[100]$ direction, the transition at $T_{\rm s2}$ is expected to become a crossover, which agrees with the disappearance of the Bragg peak splitting at $T_{\rm s2}$ for $H \geq 2$~T~\cite{Matsumura2022-ky}. Thus, although our scenario differs from earlier interpretations, it remains fully consistent with existing experimental observations. 

For future experiments, the presence of a phase transition at $T_{\rm s2} = 8$~K and the disappearance of the associated anomaly under a $[100]$ magnetic field, so far detected via XRD measurements, should also be observed using other probes such as specific heat and NMR. Furthermore, existing experiments have not yet determined whether ${\bm M}^{\rm AF} \parallel {\bm Q}$ or ${\bm M}^{\rm AF} \perp {\bm Q}$ is realized. In the case of ${\bm M}^{\rm AF} \parallel {\bm Q}$, the magnetic moment possesses the $c$-axis component in the temperature range $T_{\rm s2} < T < T_{\rm N}$, which should be detectable using neutron scattering, resonant x-ray scattering, or NMR experiments. Additionally, field-induced magnetic transitions expected in the ${\bm M}^{\rm AF} \parallel {\bm Q}$ scenario should become evident in experiments with a sufficient resolution.

In summary, we found the emergence of the third zero-field phase transition and possible field-induced magnetic phases. The quadrupole order parameter acts as a weak symmetry-breaking field for the antiferromagnetic transition, resulting in three successive transitions. Although this scenario differs from previously proposed ones, it remains fully consistent with existing experimental results. The third transition should also be detectable via bulk measurements or NMR measurements. It remains unclear whether the antiferromagnetic moment possesses a finite $c$-axis component in the intermediate phase in $T_{\rm s2}<T<T_{\rm N}$. To clarify the order parameters, one could perform neutron scattering, resonant x-ray scattering, or NMR measurements within the temperature range $T_{\rm s2} < T < T_{\rm N}$. If the $c$-axis component exists, field-induced magnetic phases are expected to emerge. These results provide a basis for designing experiments to elucidate the phase diagrams and the order parameters of CeCoSi.

{\it Acknowledgement.---}This work was supported by the JSPS KAKENHI (No. JP23K20824).

%

\end{document}